%% file: main.tex
\pdfoutput=1

\makeatletter \@input{texdirectives.tex} \makeatother

\let\mypdfximage\pdfximage
\def\pdfximage{\immediate\mypdfximage}

            \documentclass[a4paper,USenglish,autoref]{lipics-v2021}
      
  \author{Jérémy Thibault}{MPI-SP, Bochum, Germany}{jeremy.thibault@mpi-sp.org}{https://orcid.org/0009-0008-5112-3269}{}
  \author{Joseph Lenormand}{MPI-SP, Bochum, Germany \and ENS Paris-Saclay, France}{joseph.lenormand@ens-paris-saclay.fr}{}{}
  \author{C\u{a}t\u{a}lin Hri\c{t}cu}{MPI-SP, Bochum, Germany}{catalin.hritcu@mpi-sp.org}{https://orcid.org/0000-0001-8919-8081}{}
  \authorrunning{J. Thibault, J. Lenormand, and C. Hri\c{t}cu}
  \Copyright{J. Thibault, J. Lenormand, and C. Hri\c{t}cu}
  \ccsdesc[500]{Security and privacy~Logic and verification}
  \ccsdesc[500]{Security and privacy~Formal methods and theory of security}
  \ccsdesc[300]{Software and its engineering~Compilers}
  \keywords{secure compilation;
back-translation;
machine-checked proofs;
Rocq;
Coq}
  \relatedversiondetails{arXiv Version}{https://arxiv.org/abs/2503.19609}
  \supplement{}
  \supplementdetails[subcategory={Source Code}, cite={}, swhid={}]{Software}{https://github.com/secure-compilation/nanopass-bt}

\acknowledgements{
We thank the anonymous reviewers for their helpful feedback.
This work was in part supported
by the {European Research Council}{}
under ERC{} Starting Grant SECOMP ({715753}),
and by the Deutsche Forschungsgemeinschaft (DFG, German Research Foundation)
as part of the Excellence Strategy of the German Federal and State Governments
-- EXC 2092 CASA - 390781972.}
\nolinenumbers %

\usepackage{forest}
\usepackage{mathpartir}

\usepackage{fancyvrb} %
\usepackage{listings}
\usepackage{soul}

\usepackage{xspace}

\definecolor{dkblue}{rgb}{0,0.1,0.5}
\definecolor{dkgreen}{rgb}{0,0.4,0}
\definecolor{dkred}{rgb}{0.6,0,0}
\definecolor{dkpurple}{rgb}{0.7,0,1.0}
\definecolor{purple}{rgb}{0.9,0,1.0}
\definecolor{olive}{rgb}{0.4, 0.4, 0.0}
\definecolor{teal}{rgb}{0.0,0.4,0.4}
\definecolor{azure}{rgb}{0.0, 0.5, 1.0}
\definecolor{gray}{rgb}{0.5, 0.5, 0.5}
\definecolor{dkgray}{rgb}{0.3, 0.3, 0.3}

\usepackage{xargs}

\newcommand{\RSP}{RSP\xspace}
\newcommand{\RFrSP}{RFrSP\xspace}
\newcommand{\prg}[1]{\mathord{\mathtt{#1}}}

\newcommandx{\cmp}[3][2,3]{\mathord{\ensuremath{#1\!\!\downarrow^{#2}_{#3}}}}
\newcommand{\back}[1]{\ensuremath{#1\!\!\uparrow}}

\newcommand{\red}[1]{\textcolor{red}{#1}}
\newcommand{\ecall}[3]{\mathrm{Call}~#1~#2~#3}
\newcommand{\eret}[3]{\mathrm{Ret}~#1~#2~#3}

\newcommand{\loc}{\mathtt{loc}}

\newcommand{\tree}[1]{\mathord{\mathsf{#1}}}

\newcommand{\loca}[1]{\red{#1}}
\newcommand{\node}[3]{\loca{#1} : #2 : \loca{#3}}

\newcommand*{\EG}{e.g.,\xspace}
\newcommand*{\IE}{i.e.,\xspace}
\newcommand*{\ETAL}{et al.\xspace}
\newcommand*{\ETC}{etc.\xspace}

\newcommand{\comm}[3]{}

\usetikzlibrary{positioning}

\def\Snospace~{\S{}}

\def\Nnospace~{}

\definecolor{ltblue}{rgb}{0,0.4,0.4}
\definecolor{dkblue}{rgb}{0,0.1,0.6}
\definecolor{dkgreen}{rgb}{0,0.35,0}
\definecolor{dkviolet}{rgb}{0.3,0,0.5}
\definecolor{dkred}{rgb}{0.5,0,0}
\lstdefinelanguage{Rocq}{
    mathescape=true,
    texcl=false,
    morekeywords=[1]{Section, Module, End, Require, Import, Export,
        Variable, Variables, Parameter, Parameters, Axiom, Hypothesis,
        Hypotheses, Notation, Local, Tactic, Reserved, Scope, Open, Close,
        Bind, Delimit, Definition, Let, Ltac, Fixpoint, CoFixpoint, Add,
        Morphism, Relation, Implicit, Arguments, Unset, Contextual,
        Strict, Prenex, Implicits, Inductive, CoInductive, Record,
        Structure, Canonical, Coercion, Context, Class, Global, Instance,
        Program, Infix, Theorem, Lemma, Corollary, Proposition, Fact,
        Remark, Example, Proof, Goal, Save, Qed, Defined, Hint, Resolve,
        Rewrite, View, Search, Show, Print, Printing, All, Eval, Check,
        Projections, inside, outside, Def},
    morekeywords=[2]{forall, exists, exists2, fun, fix, cofix, struct,
        match, with, end, as, in, return, let, if, is, then, else, for, of,
        nosimpl, when},
    morekeywords=[3]{Type, Prop, Set, true, false, option},
    morekeywords=[4]{pose, set, move, case, elim, apply, clear, hnf,
        intro, intros, generalize, rename, pattern, after, destruct,
        induction, using, refine, inversion, injection, rewrite, congr,
        unlock, compute, ring, field, fourier, replace, fold, unfold,
        change, cutrewrite, simpl, have, suff, wlog, suffices, without,
        loss, nat_norm, assert, cut, trivial, revert, bool_congr, nat_congr,
        symmetry, transitivity, auto, split, left, right, autorewrite},
    morekeywords=[5]{by, done, exact, reflexivity, tauto, romega, omega,
        assumption, solve, contradiction, discriminate},
    morekeywords=[6]{do, last, first, try, idtac, repeat},
    morecomment=[s]{(*}{*)},
    showstringspaces=false,
    morestring=[b]",
    morestring=[d]’,
    tabsize=3,
    extendedchars=false,
    sensitive=true,
    breaklines=false,
    basicstyle=\small,
    columns=[l]flexible,
    identifierstyle={\ttfamily\color{black}},
    keywordstyle=[1]{\ttfamily\color{dkviolet}},
    keywordstyle=[2]{\ttfamily\color{dkgreen}},
    keywordstyle=[3]{\ttfamily\color{ltblue}},
    keywordstyle=[4]{\ttfamily\color{dkblue}},
    keywordstyle=[5]{\ttfamily\color{dkred}},
    stringstyle=\ttfamily,
    commentstyle={\ttfamily\color{dkgreen}},
    literate=
    {\\forall}{{\color{dkgreen}{$\forall\;$}}}1
    {\\exists}{{$\exists\;$}}1
    {<-}{{$\leftarrow\;$}}1
    {=>}{{$\Rightarrow\;$}}1
    {<>}{{\(\neq\;\)}}1
    {==}{{\(=\;\)}}1
    {->}{{$\rightarrow\;$}}1
    {<->}{{$\leftrightarrow\;$}}1
    {<==}{{$\leq\;$}}1
    {\#}{{$^\star$}}1
    {\\o}{{$\circ\;$}}1
    {\@}{{$\cdot$}}1
    {\/\\}{{$\wedge\;$}}1
    {\\\/}{{$\vee\;$}}1
    {~}{{$\sim$}}1
    {\@\@}{{$@$}}1
    {\\mapsto}{{$\mapsto\;$}}1
    {\\hline}{{\rule{\linewidth}{0.5pt}}}1
}[keywords,comments,strings]

\newcommand{\papertitle}{Nanopass Back-Translation of Call-Return Trees for Mechanized Secure Compilation Proofs}

\makeatletter
\AtBeginDocument{
  \hypersetup{
    pdftitle = {\papertitle},
    pdfauthor = {Jérémy Thibault, Joseph Lenormand, and Cătălin Hrițcu}
  }
}
\makeatother

\renewcommand{\paragraph}[1]{{\bf #1.}\;}

\begin{document}

\title{\papertitle}
\titlerunning{Nanopass Back-Translation of Trees for Secure Compilation}
\maketitle

\begin{abstract}
  \input{abstract.txt}
\end{abstract}
\section{Introduction}

Good programming languages provide useful abstractions for writing more secure
code, such as procedures, %
types, modules, interfaces,
or even something as basic as structured control flow.
However, compiling and linking with untrusted low-level code---for instance with a
library written in an unsafe language like C/C++ or
assembly---doesn't necessarily preserve the security properties of the source program.
Indeed, such linked low-level code---which we call a target \emph{context}---doesn't have
to respect the same abstractions as source contexts, and, if malicious or compromised, can
actively attack the compiled program.

Many prototype \emph{secure compilation chains}~\cite{AbateABEFHLPST18,
  MarcosSurvey, DevriesePPK17, difftraces, AbateBGHPT19, AbadiP12,
  PatrignaniASJCP15, FournetSCDSL13, AgtenSJP12, NewBA16, PatrignaniG21,
  StrydonckPD19, ThibaultBLAAGHT24, secureptrs, GeorgesTB22}
aim to protect a compiled program from such attacks.
More precisely, they enforce that if there is no attack a source context
can mount against a source program, then there can be no attack an adversarial
target context can mount against the compiled program.
An attack can be represented as one or multiple execution traces,
often also recording the interactions between program and context.
Such an attack is a witness of the violation of a security property, which can be
a trace property (\EG safety), a hyperproperty~\cite{ClarksonS10}
(\EG noninterference~\cite{GoguenM82}), or a relational hyperproperty~\cite{AbateBGHPT19}
(\EG trace equivalence, obtaining a well-investigated
criterion called full abstraction~\cite{Abadi99,MarcosSurvey}).
One may be interested in protecting various such security property classes
against attacks, resulting in various formal criteria for secure
compilation~\cite{AbateBGHPT19}.

Achieving such secure compilation criteria is, however, challenging; and even more
challenging is proving formally that it satisfies any of these criteria.
Most proofs rely on a notion of {\em back-translation}~\cite{DevriesePPK17, NewBA16,
  AbateABEFHLPST18, AgtenSJP12, FournetSCDSL13, StrydonckPD19, ThibaultBLAAGHT24,
  secureptrs, DevrieseMP24}:
given a target context performing an attack on a compiled program, often evidenced by
execution traces, one must build a source context performing the
same attack against the source program.
Such back-translations and their corresponding proofs can be highly non-trivial,
especially when the attacks are against hyperproperties and relational
hyperproperties.
For such classes of security properties, attacks are witnessed by several traces
that the generated context must be able to produce, depending on the
observable behavior of the program with which it is linked.

Moreover, existing secure compilation proofs are often hard to trust, since they
can span several hundred pages when carefully spelled out even for toy
compilers~\cite{akram-capabileptrs, PatrignaniG21, DevriesePPK17} and often involve
complex reasoning techniques, such as cross-language logical
relations~\cite{NewBA16,DevriesePPK17}.
Because in practice every flaw would be potentially exploitable by the adversarial context,
it is indispensable that the trust in secure compilation proofs be very high.
So proof assistants are starting to be employed for such
proofs~\cite{AbateABEFHLPST18, DevriesePPK17, secureptrs, GeorgesTB22, ThibaultBLAAGHT24},
yet, the problem of designing general proof techniques that are easier to
mechanize in a proof assistant is still open.
Recent work in the field showed that proof assistants can help scale up secure
compilation proofs to a CompCert-based compilation chain for a compartmentalized
variant of C~\cite{ThibaultBLAAGHT24}, by allowing the proofs to be carefully
designed and engineered as if they were programs.
Yet, even this recent milestone does not address the challenge of proving secure
compilation for hyperproperties.

\paragraph{Contributions}
As an important step towards large-scale secure compilation proofs for
hyperproperties and relational hyperproperties, we propose a novel and general
technique for back-translating a finite set of finite trace prefixes into a
single source context, and a convenient way to prove the correctness of this
back-translation in a proof assistant.

We observe that the set of finite trace prefixes recording the calls and returns
between the program and the context can be represented as a \emph{call-return
  tree} branching on calls and returns from the program to the context.
During the execution, the context produced by our back-translation has to deduce,
from the events that have already been executed, which branch it is running (\IE what subset of
programs it could still be linked with),{}
in order to know what event to emit next.
To do so, the context produced by our back-translation maintains private state
storing its last \emph{location} in the tree. %
Every time the context gets control, it uses this information as well as the
argument or the return value from the program, to decide what event to produce
next, and how to update its own location.

This raises two main challenges:
(1)~the generated context must correctly handle {\em reentrancy},
since the program can perform arbitrary {\em callbacks};
and (2)~because these callbacks can be arbitrarily nested, returns may occur
arbitrarily far from their corresponding call.
So the generated context manipulates a complex state that is
difficult to reason about.
Hence, as the authors have experienced first-hand in previous failed attempts
to produce a mechanized proof, a monolithic approach leads to writing a complex
back-translation function and to dealing with many invariants at the same time
trying to prove this complex back-translation correct.
To address these challenges, we instead structure the back-translation as {\em a
nanopass compiler}~\cite{DBLP:conf/els/Keep20, SarkarWD05, KeepD13}.
The back-translation is done in several small steps, each performing a simple transformation of the tree,
adding to each node new information describing
how the location in the tree changes depending on how the context regains control.
This information is then used to generate the source context that produces the
right trace prefixes.

We prove this back-translation correct by adapting compiler correctness techniques.
We give semantics to each intermediate tree language and prove several small
CompCert-style forward simulations~\cite{Leroy09b,TanMKFON19} between these representations.
These small-step operational semantics reduce a tree by following a trace: for each
event of the trace, the semantics follow the corresponding branch, and emit the event.
The semantics also explicitly enforce invariants of the execution, using
\emph{ghost state} to store information, such as what events have already been
executed, or an abstract call stack.
In the last compilation step, the simulation proofs link this ghost state to a
\emph{concrete state} from the source language semantics.

Thanks to this modular structure, the effort of proving these small
transformations from trees to trees and eventually to source code is lower,
and we have fully mechanized these transformations and their correctness proofs
in the Rocq (formerly Coq) proof assistant~\cite{coq}.
The back-translation targets a simple source language with procedures and
compartments, and is designed to replace the existing back-translation step of
an existing formally secure compartmentalizing compilation chain~\cite{AbateABEFHLPST18}.
While we have not yet mechanized this part, it should be relatively easy to show
that this compilation chain satisfies a much stronger secure compilation
criterion---\IE that it robustly preserves a class of relational hyperproperties
including hypersafety (\EG noninterference) and relational hypersafety
properties (\EG trace equivalence, thus also obtaining full abstraction)---by
simply replacing the previous simplistic back-translation handing only one
finite trace prefix with ours that back-translates any finite set of finite
trace prefixes (see \autoref{sec:goal}).

Finally, we believe that this general technique should be reusable for other
secure compilation proofs.
In particular, the nanopass approach allows us to structure the back-translation
into two distinct phases: a tree-manipulating phase that only depends on
the trace model, but on no other language features, and a code generation phase.
Even the code generation phase and the proofs pertaining to it should be
mostly reusable, as they only rely on programming language features that are
common to many languages (\EG procedures, private state).
For this reason, we see this work as an important step in devising secure
compilation proof techniques that can be scaled up to real-world languages,
such as the recent compartmentalized version~\cite{ThibaultBLAAGHT24} of CompCert
(see \autoref{sec:future}).

The remainder of the paper is structured as follows: \autoref{sec:main} first
gives a higher-level account of our back-translation and proof technique,
explaining the key ideas.
\autoref{sec:tree-langs} then presents the design of the semantics of the
intermediate languages and of the back-translation passes.
\autoref{sec:proof-sim} then outlines how we proved correctness for this
back-translation using forward simulations.
Afterwards, \autoref{sec:implementation} describes our Rocq formalization,
\autoref{sec:related} discusses related work, and \autoref{sec:future} closes
with future work.

Our Rocq formalization is available at \url{https://github.com/secure-compilation/nanopass-bt}.
\section{Background: secure compilation of compartmentalized code}\label{sec:background}

Abate \ETAL~\cite{AbateABEFHLPST18} showed how one can define a secure
compilation criterion for a setting with mutually-distrustful compartments and
undefined behavior. They reduce this criterion to a simpler,
trace-based criterion, a variant of \emph{Robust Safety Preservation}
(\RSP{})~\cite{AbateBGHPT19, PatrignaniG21} and verify that a simple compiler
satisfies this criterion in Rocq.
Their proof technique is of particular interest to us, because a more recent
work by Thibault \ETAL~\cite{ThibaultBLAAGHT24} reused it at scale for a compartmentalized
variant of the C language and a realistic compiler based on CompCert.
By strengthening the back-translation step, it is easy to extend this proof
technique to show a much stronger secure compilation criterion, as illustrated
informally in the next subsection.
But first, we present the setting of Abate \ETAL~\cite{AbateABEFHLPST18}, which we also use in this paper.

In this setting, programs consist in several mutually-distrustful compartments \(\prg{C_1}, \prg{C_2}, \dots\) that interact
with each other via arbitrary procedure calls and returns, and can pass integers to each other.
To simplify the exposition throughout the paper,
we make the simplifying assumption that each compartment is made of only one procedure, and we equate the compartments and the procedure each contains.
{}%
Cross-compartment interactions happen via calls and returns and are recorded on
a trace; for instance, \(m_1\) is a finite trace prefix capturing
some calls and returns between two compartments \(\prg{C_1}\)
and \(\prg{C_2}\):
\(
  m_1~=~\ecall{\prg{C_1}}{\prg{C_2}}{40};
        \ecall{\prg{C_2}}{\prg{C_1}}{41};
        \eret{\prg{C_1}}{\prg{C_2}}{42};
        \eret{\prg{C_2}}{\prg{C_1}}{42}
\).

Abate \ETAL~\cite{AbateBGHPT19} define a hierarchy of secure compilation criteria that are
based on such execution traces, among which \emph{Robust Safety Preservation}
(\RSP{}), one of the simplest criteria, is the one proved
by Abate \ETAL~\cite{AbateABEFHLPST18}:\footnote{In fact they prove a slight variant of
  \RSP~\cite{difftraces}, but for simplicity we stick with standard \RSP here.}
\begin{definition}\label{def:rsp}
A compilation chain is said to satisfy \RSP{} when:
\[
  \forall\mathbb{P}_{\mathtt{S}}~\mathbb{C}_{\mathtt{T}}~m\ldotp~
      \cmp{\mathbb{P}_{\mathtt{S}}} \mathbin{\cup} \mathbb{C}_{\mathtt{T}} \rightsquigarrow m
  \Rightarrow
  \exists \mathbb{C}_{\mathtt{S}}\ldotp~
      \mathbb{P}_{\mathtt{S}} \cup \mathbb{C}_{\mathtt{S}} \rightsquigarrow m
\]
\end{definition}
\noindent
where \(\mathbb{C}_{\mathtt{S}}\) and \(\mathbb{C}_{\mathtt{T}}\)
are contexts, and \(\mathbb{P}_{\mathtt{S}}\)  a partial
program, which are all sets of compartments, \(\cup\) represents
linking, \(\cmp{\mathbb{P}}\) denotes the compilation of \(\mathbb{P}\),
\(m\) is a finite trace, and \(\rightsquigarrow\) represents the trace-producing semantics.%
\footnote{Objects are annotated with \(\mathtt{S}\) for the source language and
with \(\mathtt{T}\) for the target language whenever it is necessary to
disambiguate. We sometimes omit the \(\mathtt{S}\) annotation to simplify notations.}
Informally, \RSP{} means the following: for any target context \(\mathbb{C}_{\mathtt{T}}\) that
mounts an attack against a compiled (partial) program
\(\cmp{\mathbb{P}_{\mathtt{S}}}\) characterized by the finite trace prefix \(m\), one can show
that the attack \(m\) can already be mounted against the program \(\mathbb{P}_{\mathtt{S}}\)
in the source language, \IE{}
there exists a source context \(\mathbb{C}_{\mathtt{S}}\) that can be linked with \(\mathbb{P}_{\mathtt{S}}\)
to produce the same finite trace prefix \(m\).
In this setting both the context and the partial program
are (disjoint) sets of compartments that can be put together by the linking
operation \(\cup\) to produce a whole program, and they only differ in the role
they play in the theorem: the program is securely compiled and protected from adversarial target contexts.
\RSP{} as well as the stronger criterion below are general properties
of a whole compilation chain, including not only the compiler, but also the
source and target linkers as well as the source and target semantics, which are
all jointly responsible for achieving security.

Abate \ETAL~\cite{AbateABEFHLPST18} proved that their compilation chain satisfies \RSP{},
via a proof technique that uses a trace-based back-translation.
The core idea behind their technique can be found in \autoref{fig:proof-technique},
by ignoring the parts in red (or equivalently setting \(K = 1\)).
In summary, they define a back-translation function \(\uparrow\) that, from the
trace \(m\), generates a whole program (\(\mathbb{P}_{\mathtt{S}}' \cup \mathbb{C}_{\mathtt{S}}\))
that produces the same trace \(m\).
This program is compiled, and via a step called recomposition, they prove
that a recomposed program \(\cmp{\mathbb{P}_{\mathtt{S}}}
 \cup \cmp{\mathbb{C}_{\mathtt{S}}} \rightsquigarrow m\).
They finally use backward compiler correctness to go back to the source level, obtaining:
\(\mathbb{P}_{\mathtt{S}} \cup \mathbb{C}_{\mathtt{S}} \rightsquigarrow m\).
To use compiler correctness in such manner, Abate \ETAL~\cite{AbateABEFHLPST18} assume
separate compilation~\cite{KangKHDV15}, which can be summarized as linking and
compilation commuting.

One interesting characteristic of the back-translation in this setting is that
it produces \emph{whole programs} instead of just a context.
Abate \ETAL~\cite{AbateABEFHLPST18} assert that this considerably simplifies proving the back-translation correct, since it
allows them to reason operationally about the behavior of a whole program
produced entirely by their back-translation.
If instead the back-translation produced only the source context, they would
need either to reason about its behavior when linked with arbitrary source
programs, or to introduce a partial semantics (often called ``trace
semantics''~\cite{PatrignaniC15, BusiNBGDMP20, akram-capabileptrs, AgtenSJP12})
that uses nondeterminism to abstract the missing program part.
These options are, however, more involved than simply working with whole
programs.
Working with whole programs, moreover, enables the use of a standard notion of
compiler correctness (with separate compilation, like for current
CompCert~\cite{KangKHDV15}) to go up and down between the source and target
languages (steps II and IV in \autoref{fig:proof-technique}).

Note also that this back-translation constructs the source context \(\mathbb{C}_{\mathtt{S}}\)
of \autoref{def:rsp} using solely the information that appears on \(m\) (as
opposed to inspecting the target context
\(\mathbb{C}_{\mathtt{T}}\)~\cite{FournetSCDSL13, NewBA16, DevriesePPK17, StrydonckPD19}; see
\autoref{sec:related} for a comparison).
Abate \ETAL~\cite{AbateABEFHLPST18} show how to build a source context that faithfully
reproduces such a single trace using a very simple scheme, also used in all the
other trace-based back-translations~\cite{PatrignaniC15, BusiNBGDMP20,
  akram-capabileptrs, AgtenSJP12}.
To understand how this back-translation is implemented, let us look at the
back-translation of the previously defined \(m_1\) for \(\prg{C_1}\):
\(
  m_1 = \ecall{\prg{C_1}}{\prg{C_2}}{40};
        \ecall{\prg{C_2}}{\prg{C_1}}{41};
        \eret{\prg{C_1}}{\prg{C_2}}{42};
        \eret{\prg{C_2}}{\prg{C_1}}{42}.
\)
\begin{lstlisting}[numbers=left, basicstyle=\ttfamily\footnotesize, columns=fullflexible, keepspaces=true,xleftmargin=1.5\parindent,caption={Single trace back-translation of \(m_1\)},label=lst:rsc-bt,frame=single]
C1():
if (ctr = 0) {        // first time called
  ctr++;              // update counter
  C2(40);             // perform first event
  C1();              // recursive call
} else if (ctr = 1) { // second time called
  ctr++;              // update counter
  return 42;          // perform second event
}
\end{lstlisting}

The context maintains a private counter {\tt ctr}, storing how many events have
already been produced.
When \(n{-}1\) events have been produced and the context gets back control it
simply produces the \(n\)-th event of the trace.
The recursive call to {\tt C1} is an internal call that
doesn't appear on the trace, and allows performing several events sequentially.
This back-translation is applied pointwise to each compartment, so as to obtain
the full back-translation.

\section{Key Ideas}\label{sec:main}

\subsection{Towards a proof of a stronger criterion}\label{sec:goal}
While Abate \ETAL~\cite{AbateABEFHLPST18} proved that their compilation chain satisfies \RSP{},
this is still a weak secure compilation criterion, one of the weakest they study.
Indeed, \RSP{} only considers attacks represented by a single finite trace prefix.
What this means is that a secure compilation chain satisfying \RSP{} only
robustly preserves safety properties.
In general, safety properties cannot capture data confidentiality and integrity, which are essential security properties that one would want their compilation chain to preserve.

Thankfully, Abate \ETAL~\cite{AbateBGHPT19} also propose stronger criteria that consider several trace
prefixes, which can also represent attacks against data confidentiality and integrity.
One such criterion is the one named \emph{Robust
Finite-relational Safety Preservation} (\RFrSP):
\begin{definition}\label{def:rfrsp}
A compilation chain is said to satisfy \RFrSP{} when:
\begin{align*}
  &\forall K~
    \forall\mathbb{P}_1\dots\mathbb{P}_K~
    \mathbb{C}_{\mathtt{T}}~m_1\dots m_K.~\\
  &\quad (\cmp{\mathbb{P}_1} \cup \mathbb{C}_{\mathtt{T}}
  \rightsquigarrow m_1
  \mathrel{\wedge} \dots \mathrel{\wedge}
  \cmp{\mathbb{P}_K} \cup \mathbb{C}_{\mathtt{T}}
  \rightsquigarrow m_K)
  \implies\\
  &\quad \exists \mathbb{C}_{\mathtt{S}}\ldotp
  (\mathbb{P}_1 \cup \mathbb{C}_{\mathtt{S}} \rightsquigarrow m_1
  \mathrel{\wedge} \dots \mathrel{\wedge}
  \mathbb{P}_K \cup \mathbb{C}_{\mathtt{S}} \rightsquigarrow m_K).
\end{align*}
\end{definition}
\RFrSP{} states that any attack against a finite-relational safety property
of compiled programs $\mathbb{P}_1\dots\mathbb{P}_K$ characterized by the finite
trace prefixes $m_1\dots m_K$, can also be mounted against the source programs
(\IE{} there exists a source context \(\mathbb{C}_{\mathtt{S}}\) that induces the same prefixes).

This criterion is particularly interesting, as it implies the robust
preservation of all hypersafety properties~\cite{ClarksonS10}, a large class
that includes not just safety but also most variants of noninterference.
In our context, it also implies a variant of (the security-relevant direction
of) the well-studied criterion of fully abstract compilation~\cite{Abadi99,
  AbateBGHPT19}.
Finally, this is the strongest criterion of Abate \ETAL~\cite{AbateBGHPT19} that can be
proved by back-translating trace prefixes.

The \RFrSP{} criterion is, however, harder to prove than \RSP{}: while for \RSP{}
the context generated by back-translation only needs to produce a single trace
prefix when linked with a single known program, for \RFrSP{} the produced
context needs to be able to produce finitely many trace prefixes when linked
with several possibly different programs.

The first insight is that it is possible to prove \RFrSP{} using the same proof
technique, with the sole difference in the back-translation.
Indeed, suppose one is able to provide a back-translation
\(\back{(m_1,\dots,m_K)}~ = (\mathbb{C}_{\mathtt{S}},
\mathbb{P}_1',\dots, \mathbb{P}_K')\)
such that
    \(\forall i.~ \mathbb{P}_i' \cup \mathbb{C}_{\mathtt{S}} \rightsquigarrow m_i\).
Then, it is possible to plug this back-translation in place of the
single-trace back-translation and apply all the other steps pointwise.
One then obtains a proof of \RFrSP{}.
The updated proof technique is given in \autoref{fig:proof-technique},
with the differences highlighted in red.

\begin{figure} %
\centering
\input{img/technique}
\caption{Outline of Abate \ETAL~\cite{AbateABEFHLPST18}'s proof technique adapted for \RFrSP{}
  (parts changed in \red{red})}\label{fig:proof-technique}
\vspace{-1.2em}
\end{figure}

\subsection{Back-translating multiple traces}\label{sec:back-by-example}

How can one back-translate a finite set of finite trace prefixes though?
Consider the example from \autoref{sec:background}, but with compartment
\(\prg{C}_{\mathbb{C}}\) in the context and compartment \(\prg{C}_{\mathbb{P}}\)
in the program (instead of \(\prg{C_1}\) and \(\prg{C_2}\)).
Suppose that the context is linked with three other programs so that we obtain
the traces \(m_1, m_2, m_3\) on the right of \autoref{fig:ex-tree}.
A key idea is that such prefixes can be represented as a call-return tree whose
branches are labeled with events, as seen on the left of \autoref{fig:ex-tree}.
In the languages we consider branching can only occur when the program has
control, and two different programs decide to produce different events.
\begin{figure}[t!]
\begin{center}
\begin{forest}
for tree={grow'=east,parent anchor=east,child anchor=west,calign=center,l sep=5.5em, s sep=3.5em}
[\footnotesize\(0\)
 [\footnotesize\(1\),
 edge label={node[midway,auto]{\scalebox{0.7}{\(\ecall{\prg{C}_{\mathbb{C}}}{\prg{C}_{\mathbb{P}}}{40}\)}}},
  [\footnotesize\(2\),
  edge label={node[midway,auto]{\scalebox{0.7}{\(\ecall{\prg{C}_{\mathbb{P}}}{\prg{C}_{\mathbb{C}}}{41}\)}}},
   [\footnotesize\(3\),
   edge label={node[midway,auto]{\scalebox{0.7}{\(\eret{\prg{C}_{\mathbb{C}}}{\prg{C}_{\mathbb{P}}}{42}\)}}}
    [\footnotesize\(4\),
    edge label={node[midway,auto]{\scalebox{0.7}{\(\eret{\prg{C}_{\mathbb{P}}}{\prg{C}_{\mathbb{C}}}{43}\)}}},
    ]
   ]
  ]
  [\footnotesize\(5\),
  edge label={node[midway,auto]{\scalebox{0.7}{\quad~~\(\eret{\prg{C}_{\mathbb{P}}}{\prg{C}_{\mathbb{C}}}{43}\)}}},
  ]
  [\footnotesize\(6\),
  edge label={node[midway,auto]{\scalebox{0.7}{\(\eret{\prg{C}_{\mathbb{P}}}{\prg{C}_{\mathbb{C}}}{44}\)}}},
  ]
 ]
]
\node at (9, 0) [rectangle,draw] {%
\scalebox{0.7}{\parbox{0.5\textwidth}{
\text{{\bfseries Set of trace prefixes:}}
\begin{align*}
  m_1 &= \ecall{\prg{C}_{\mathbb{C}}}{\prg{C}_{\mathbb{P}}}{40};
        \ecall{\prg{C}_{\mathbb{P}}}{\prg{C}_{\mathbb{C}}}{41};
        \eret{\prg{C}_{\mathbb{C}}}{\prg{C}_{\mathbb{P}}}{42};
        \eret{\prg{C}_{\mathbb{P}}}{\prg{C}_{\mathbb{C}}}{42}\\
  m_2 &= \ecall{\prg{C}_{\mathbb{C}}}{\prg{C}_{\mathbb{P}}}{40}; \eret{\prg{C}_{\mathbb{P}}}{\prg{C}_{\mathbb{C}}}{43}\\
  m_3 &= \ecall{\prg{C}_{\mathbb{C}}}{\prg{C}_{\mathbb{P}}}{40}; \eret{\prg{C}_{\mathbb{P}}}{\prg{C}_{\mathbb{C}}}{44}
\end{align*}
}}
};
\end{forest}
\end{center}
\vspace{-1em}
\begin{lstlisting}[numbers=left, basicstyle=\ttfamily\scriptsize, columns=fullflexible, keepspaces=true,xleftmargin=1.5\parindent,frame=single]
(*@\(\prg{C}_{\mathbb{C}}\)@*)(arg):
if (is_call) {
  /* Receiving a cross-compartment callback; we update the location based
     on the current location and the received argument */
  if (loc = 1) {
    if (arg = 41)
      loc = 2;
  }
} else {
  /* this is not an incoming call, but rather a recursive call to handle
     a return value. We update the location based on the current location
     and the returned value (stored in res) */
  if (loc = 1) {
    if (res = 43)
      loc = 5;
    else if (res = 44)
      loc = 6;
  } else if (loc = 3) {
    if (res = 43)
      loc = 4;
  } };
// Now we generate the appropriate event based on the current location
if (loc = 0) {
  is_call = 1;
  loc = 1;
  res = (*@\(\prg{C}_{\mathbb{P}}\)@*)(40);
  is_call = 0;
  (*@\(\prg{C}_{\mathbb{C}}\)@*)();
} else if (loc = 2) {
  is_call = 1;
  loc = 3;
  return 42; }
\end{lstlisting}
\caption{Example of a call-return tree representing a set of finite prefixes, and the
  corresponding code for \(\prg{C}_{\mathbb{C}}\) generated by the
  back-translation}\label{fig:ex-tree}
\vspace{-1.5em}
\end{figure}

In \autoref{fig:ex-tree}, we give each node a unique identifier.
These identifiers generalize the scheme of Abate \ETAL~\cite{AbateABEFHLPST18}: instead of
relying on a counter that is always incremented, the context is now able to
encode what events have already been executed by simply recording the \emph{location}
of the last executed event in a private variable \(\loc\).
The compartment being back-translated can then use this information to determine
which event to perform next and how to update \(\loc\).
Determining how to update the \(\loc\) counter and which event to
produce is, however, far more involved than in the simple single-trace
back-translation.
In the lower part of \autoref{fig:ex-tree} we illustrate what compartment
\(\prg{C}_{\mathbb{C}}\) we generate from the traces in \autoref{fig:ex-tree}.
The core idea of the back-translation is for the
generated compartment \(\prg{C}_{\mathbb{C}}\) to maintain a private variable \(\mathtt{is\_call}\) that will be set to \(1\)
whenever \(\prg{C}_{\mathbb{C}}\) is about to pass control to another compartment (and generally when it doesn't have control).
That way, it knows that a call it receives when \(\mathtt{is\_call} = 1\) comes from another compartment (line 2), and that it must update its location \(\loc\) based on the current value
\(\loc\) and the argument it received (lines 5--7).

Following the code in \autoref{fig:ex-tree}, \(\prg{C}_{\mathbb{C}}\) goes directly to line 23, and starts branching on its
current location (that just got updated).
Depending on this value, it will produce either a cross-compartment call (lines 23--28) or a cross-compartment return (lines 29--33).
In the case of a call, the compartment sets \(\mathtt{is\_call}\) to \(1\) (line 24) and
updates its location (line 25), before calling the other compartment (line 26).
When this new call returns, the return value will be stored in a third private
variable \(\mathtt{res}\) (line 26), the compartment will set \(\mathtt{is\_call}\) to \(0\) (line 27),
and call itself recursively (line 28).
Then, the \(\mathtt{else}\) path of the first \(\mathtt{if}\) will be taken (line 9), and the return value will be used, together with \(\loc\), to set a new location (lines 13--21).
One may wonder why not set the new location right after the call, between lines 26 and 28?
This is because \(\prg{C}_{\mathbb{C}}\) might have received callbacks that updated \(\loc\)
in the meantime.
Because \(\loc\) might have changed, it is necessary for correctness to branch on it again;
and it is simpler in terms of reasoning to gather all this branching at the same place
in the code instead of it being dispersed among all the calls.

Finally, what happens when doing a return (lines 30--32)? Because it is about to pass control back,
\(\prg{C}_{\mathbb{C}}\) sets \(\mathtt{is\_call}\) to \(1\), updates its location, and starts returning.
Then, the return pops the stack and reaches another compartment,
which if also generated by our back-translation, will handle its return value
in the way described above (lines 24--29).

Note that in our example, we only consider two compartments.
In the general case, we support an arbitrary number of distinct compartments;
we consider one tree per compartment: for a
compartment \(\prg{C}\), it is the tree obtained by filtering out all events
that do not involve \(\prg{C}\) from the traces.
We are able to do this filtering because a compartment cannot observe the interactions
between two other compartments directly.
This generalization does not affect the structure of the back-translated programs.
\subsection{Nanopass back-translation}\label{sec:nano-bt}

The back-translation described above is complex; to tame this complexity,
we divide the back-translation into a series of small steps, each improving
upon the previous one in a minimal way, in the style of a nanopass
compiler~\cite{DBLP:conf/els/Keep20, SarkarWD05, KeepD13}.

For this we define several {\em intermediary tree languages} with
more and more detailed semantics as we progress through the back-translation.
Each step makes explicit one {\em invariant} that was previously implicit,
or introduces a new piece of {\em ghost state} that will later be related
to concrete state in the generated source code.

\begin{figure*}
\centering
\begin{tikzpicture}
  \node[draw,rectangle,minimum height=3em,minimum width=8em,align=center](int-prg)
  at (0,0)
  {\footnotesize Target language:\\
   \footnotesize
  \(\prg{C_T}, \cmp{\prg{P}_1}, \dots, \cmp{\prg{P}_K}\)};

  \node[draw,ellipse,minimum height=3em,minimum width=10.5em,align=center](traces)
  at (5,0)
  {\footnotesize (0) Well-formed traces\\
   \footnotesize \(m_1, \dots, m_K\)};

  \node[draw,rectangle,minimum height=3em,minimum width=10.5em,align=center](simple-trees)
  at (10,0)
  {\footnotesize (1) Call-return trees};

  \node[draw,rectangle,minimum height=3em,minimum width=10.5em,align=center](id-trees)
  at (10, 2)
  {\footnotesize (2) Trees \\
   \footnotesize with unique identifiers};

  \node[draw,rectangle,minimum height=3em,minimum width=10.5em,align=center](stack-trees)
  at (10, 4)
  {\footnotesize (3) Stack-aware trees};

  \node[draw,rectangle,minimum height=3em,minimum width=10.5em,align=center](flattened)
  at (10, 6)
  {\footnotesize (4) Flattened \\
   \footnotesize representation};

  \node[draw,rectangle,minimum height=3em,minimum width=8em,align=center](source)
  at (0, 6)
  {\footnotesize (5) Source language:\\
   \footnotesize \(\prg{C_S}, \prg{P}'_1, \dots, \prg{P}'_K\)};

  \draw[->] (int-prg) to node {} (traces);
  \draw[->] (traces) to node[above] {(a)} (simple-trees);
  \draw[->] (simple-trees) to node[left] {(b)} (id-trees);
  \draw[->] (id-trees) to node[left] {(c)} (stack-trees);
  \draw[->] (stack-trees) to node[left] {(d)} (flattened);
  \draw[->] (flattened) to node[above] {(e)} (source);
  \draw[->,double,dashed] (int-prg) to node {} (source);

\end{tikzpicture}
\caption{Steps of our nanopass back-translation}\label{fig:nanopass-bt}
\vspace{-1.5em}
\end{figure*}

\autoref{fig:nanopass-bt} gives an overview of the different passes of our back-translation.
Starting from the target language on the bottom left, the compiled programs produce a well-formed set of traces
when linked with the target context.{}
Our notion of well-formedness captures the notion of determinacy necessary for them to be
representable as trees.
This well-formed set of traces is the starting point of our formally verified back-translation:
\begin{alphaenumerate}
\item we prove they can be represented as call-return trees, as in \autoref{fig:ex-tree}
\item to each node of the trees we assign a unique identifier, which will be
  mapped to the variable \(\loc\) in the code;
\item we add an abstract cross-compartment call stack
  to the semantics of the trees;
\item we flatten the trees into lists of rules that describe how each compartment should update its private state and emit events, depending on its current private state and the last event that was executed.
For instance, a rule of the form \((n, \ecall{\prg{C_1}}{\prg{C_2}}{40}, n')\)
can mean---depending on whether it's a rule for \(\prg{C_1}\) or \(\prg{C_2}\)---either
that when at location \(n\), compartment \(\prg{C_1}\) must update
its location to \(n'\) and call \(\prg{C_2}\) with argument \(40\), or that when at location \(n\), if compartment \(\prg{C_2}\) receives a call with argument \(40\) then it must update its location to \(n'\);
\item for each such rule, we generate a source statement;
for instance, the previous rule would be translated either to \lstinline[basicstyle=\ttfamily\footnotesize]{loc = n'; C2(40); ...} (for \(\prg{C_1}\)) or to \lstinline[basicstyle=\ttfamily\footnotesize]|if(loc = n && arg = 40){loc = n'}| (for \(\prg{C_2}\)).
We then assemble the final source programs from these sub-statements.
\end{alphaenumerate}

We write \(\uparrow_{\mathrm{(a-e)}}\) to refer to the whole back-translation,
and \(\uparrow_{\mathrm{(a)}}, \dots, \uparrow_{\mathrm{(e)}}\) for each back-translation step.
We see verifying this back-translation as proving \emph{compiler correctness}.
We first equip the call-return trees with simple semantics of the form
\(\tree{T} \rightsquigarrow_{n} m\) meaning that the
call-return tree \(\tree{T}\) at level \(n\) produces the trace \(m\).
A step in this semantics works by plucking and emitting the event at the root
and using the event to choose one of the child branches to reduce next.%
We use this to state the main theorem, saying that for any prefix \(m\) in the set of well-formed
traces \(\tree{S}\) we consider,
the back-translation of $\tree{S}$ to a source whole
program \(\back{\tree{S}}_{\mathrm{(a-e)}}\) can produce \(m\) in the source semantics, \(\rightsquigarrow_{5}\):
\begin{theorem}[Correctness of call-return tree back-translation]\label{thm:correct-bt}
  For any well-formed set of traces \(\tree{S}\), if \(m \in \tree{S}\) then:
  $\back{\tree{S}}_{\mathrm{(a-e)}} \rightsquigarrow_{5} m$.
\end{theorem}
To formally prove this theorem in Rocq, we rely on the nanopass structure of the back-translation,
and first prove an initialization lemma and then a compiler correctness lemma for each pass of
the back-translation from levels (1)~to~(5).
\begin{lemma}[Initialization of the back-translation]\label{lemma:correct-init}
Let \(\tree{S}\) be a well-formed set of traces.
Then there exists \(\tree{T}\) a program at level $(1)$ such that for any \(m\in\tree{S}\),
\(\tree{T} \rightsquigarrow_{(1)} m\).
\end{lemma}
\begin{lemma}[Correctness of a back-translation step]\label{lemma:correct-step}
  Let \(\tree{T}\) be a program at level $(i)$ where $i>0$ and \((\alpha)\) the step from
  \((i)\) to \((i+1)\).
  If $\tree{T} \rightsquigarrow_{(i)} m$ then $\back{\tree{T}}_{(\alpha)} \rightsquigarrow_{(i+1)} m$.
\end{lemma}

We verify each pass after the initialization by proving a simulation.
Because each transformation is simple, proving each of these simulations is
considerably simpler than proving the entire back-translation correct at once;
and while this requires more upfront setup (defining each tree language), we
found this technique much easier overall.

Proving these lemmas raises two questions that have to be answered together:
\begin{romanenumerate}
\item How do we define the semantics of each intermediate level?
\item How can we prove that the final source program obtained by back-translation
  is correct?
  In particular, what kind of \emph{invariants} are needed in this proof?
\end{romanenumerate}

First, we identify the core invariants that must hold of the generated source
code: (i) the location stored in the private state of each compartment must
always be up-to-date; and (ii) completeness properties: all events on
the trace have a corresponding statement in the back-translated code, all
events have a corresponding statement that ``consumes'' it and updates
the state, and all these statements should be reachable when in the right state.

Then, we build up incrementally to the source level by introducing these new
invariants in the semantics, little by little.
More concretely, the small-step operational semantics of our languages enforce
these invariants by adding them as requirements of the small-step transitions.
For instance, when we add the abstract cross-compartment stack, the semantics only
allows returns if they are indeed returns from and to the compartments recorded on that stack.
This enforcement is only a tool for the proof: indeed, we later prove forward
simulations between the languages, showing that these invariants do not really
restrict the behaviors of programs.
Each of these small additions incurs a proof burden, but instead of having to
deal with them all at once, our structure allows us to do it one by one, in
isolation.
Once an invariant has been established in the semantics of one level, the fact that each step is
small makes it easy to prove that it is preserved up to the source level.
Once we reach the final pass, we only have to reason about the
semantics of the source language, but all of the more complex invariants have
already been dealt with in earlier passes.
Moreover, all the levels except the very last one are pretty much independent of
the source language we target, so we expect one would be able to reuse these
parts of the proof fairly easily (\autoref{sec:future}).

In the following sections, we describe in more detail our methodology to build
the semantics of the languages and how we use them in the proofs.

\section{Design of the nanopass back-translation}\label{sec:tree-langs}
\subsection{The tree languages in the back-translation}\label{sec:tree-langs-struct}

\paragraph{Structure and invariants} The idea at the core of our definition of
all tree languages is the following: to each compartment, we associate one tree
representing the set of traces it produces.
The type \(\mathtt{tree}\) we use to represent all of our trees is given below in \autoref{lst:tree-type}.
It is parameterized by the type \(\mathtt{A}\), which differs between each tree language and
represents the data stored in each node.
The branches are labeled by \(\mathtt{event}\)s.
\begin{lstlisting}[numbers=left,language=Rocq,keepspaces=true,basicstyle=\footnotesize,
xleftmargin=1.5\parindent,
caption={Basic type used to represent trees},label=lst:tree-type]
Inductive tree (A : Type) : Type :=
| node : (* content of the node *) A ->
         (* children *) branches A -> tree A
with branches (A : Type) : Type :=
| Bnil : branches A
| Bcons : (* label *) event ->
          (* tree corresponding to that label *) tree A ->
          (* rest of the children *) branches A -> branches A
\end{lstlisting}

There is an important asymmetry in this setting: while compartments \(\prg{C} \in \mathbb{C}\) (in the context) depend on the entire set of
traces, compartments \(\prg{C} \in \mathbb{P}_i\) (in the program) only
depend on one of the trace, \(m_i\).
What this means is that, for compartments in the program, the tree that is associated to each compartment is entirely linear, and is a simple embedding of the trace in the type of tree (\IE{} they look like this:
\lstinline[language=Rocq]{node (Bcons e1 (node (Bcons e2 ... Bnil)) Bnil)}).
Conversely, for compartments in the context, the tree associated to them is branching;
yet, this branching can only occur in restricted manners.
Indeed, our languages are deterministic: this means that the only
way for a tree associated to a compartment of the context to branch
is if a compartment of the program behaved differently.

This allows us to define \emph{determinacy invariants} that we use to prove the
completeness properties discussed in the previous section.
First, we define two determinacy invariants{} that apply at all levels.
The first such invariant is \lstinline[language=Rocq]{unique_current_tree} (\autoref{lst:unique-current-tree}):
\begin{lstlisting}[numbers=left,language=Rocq,keepspaces=true,basicstyle=\footnotesize,
xleftmargin=1.5\parindent,
caption={Invariant capturing the uniqueness of the next event for the current compartment},label=lst:unique-current-tree]
Fixpoint unique_current_tree {A : Type} (C : Compartment.id) (tr : tree A) : bool :=
  match tr with
  | node _ br => unique_current_branches C br
  end
with unique_current_branches {A : Type} (C : Compartment.id) (br : branches A): bool :=
       match br with
       | Bnil => true
       | Bcons e tr br' => (unique_current_tree C tr) &&
             (if cur_comp_of_event e == C then
                match br' with | Bnil => true | _ => false end
              else unique_current_branches C br')
       end.
\end{lstlisting}
This %
invariant captures the intuition that, if a compartment has
control (line 9), then it can only execute one single next event
(captured by \lstinline[language=Rocq]{match br' with | Bnil => true | _ => false end}).
The rest of the invariant corresponds to the recursive cases: for the subtree
\lstinline[language=Rocq]{tr} (line 8) and the remaining branches \lstinline[language=Rocq]{br'} (line 11).

The second determinacy invariant, \lstinline[language=Rocq]{deterministic_tree},
is partially listed in \autoref{lst:deterministic-tree}:
\begin{lstlisting}[numbers=left,language=Rocq,keepspaces=true,basicstyle=\footnotesize,
xleftmargin=1.5\parindent,
caption={Invariant capturing the determinacy of trees},label=lst:deterministic-tree]
Inductive deterministic_tree {A : Type} : tree A -> Prop := ...
with deterministic_branches {A : Type} : branches A -> Prop := ...
| deterministic_branches_cons : forall (e : event) tr brs,
    deterministic_branches brs ->
    deterministic_tree tr ->
    not_in e brs -> (* once we find a given event, it doesn't appear in the rest *)
    deterministic_branches (Bcons e tr brs)
with not_in {A : Type} : event -> branches A -> Prop := ...
\end{lstlisting}

This determinacy invariant captures the intuition that executing
the same event always leads to the same state; \IE there are no two branches
at the same level in the tree that are labeled by the same event.
This idea can be seen in the use of a predicate
\lstinline[language=Rocq]{not_in} (line 6), which enforces that an
event cannot appear (anymore) in the rest of the children labels.

Finally, we introduce a third determinacy invariant (that we do not list here), during step (b).
In that step, we associate to each node in the trees a location identifier and
this determinacy invariant captures the uniqueness of locations.
It is very similar to \lstinline[language=Rocq]{deterministic_tree}, except that
it is about the content of the nodes instead of the labels of the branches.

\paragraph{Semantics}
As described above, programs in these languages are maps from compartments to such trees.
We give a small-step operational semantics to each of the tree languages
that works by reducing states.
As a common basis, states contain the remaining trace that is to be produced, and, for each compartment, a tree capturing the rest of the execution.
Then, the semantics are built around a reduction rule such as the one
described in \autoref{fig:basis-step}.
The idea of this rule is that one can follow the execution of a given
trace by replaying all of its events one by one, each time
reducing the trees associated with the two compartments.
In this rule, the notation \(\tree{T}_i \xrightarrow{e} \tree{T}_i'\) means that
\(\tree{T}_i'\) is a child of \(\tree{T}_i\) through a branch labeled \(e\),
and \(\mathrm{compartments}(e) = (\prg{C_i}, \prg{C_j})\) means that
the event \(e\) is a call or a return from \(\prg{C_i}\) to \(\prg{C_j}\).

\begin{figure}
\centering
\resizebox{0.9\textwidth}{!}{
\[
\inferrule[S-Step]
  { s.\mathrm{trace} = e :: t \\
    \mathrm{compartments}(e) = (\prg{C_i}, \prg{C_j}) \\
    s.\mathrm{trees}(\prg{C_i}) = \tree{T}_i \\
    s.\mathrm{trees}(\prg{C_j}) = \tree{T}_j \\
    \tree{T}_i \xrightarrow{e} \tree{T}_i' \\
    \tree{T}_j \xrightarrow{e} \tree{T}_j' \\
    s'.\mathrm{trees} = s.\mathrm{trees}[\prg{C_i} \mapsto \tree{T}_i', \prg{C_j} \mapsto \tree{T}_j'] \\
    s'.\mathrm{trace} = t
}
  { s \xrightarrow{e} s' }
\]
}
\caption{Basic reduction rule for trees}\label{fig:basis-step}
\vspace{-1em}
\end{figure}

Moreover, the semantics maintains some amount of \emph{ghost state}
\(\mathsf{gs}\), which stores information about what happened
previously during the execution.
This information contains the current location \(\mathsf{gs}.\mathrm{loc}\)
(a map from compartments to natural numbers added by step (a)),
and an abstract stack \(\mathsf{gs}.\mathrm{stack}\) (added in step (b)) recording cross compartment calls,
\IE{} a sequence of \lstinline[language=Rocq]{(Compartment.id * Compartment.id)}.
Then, the semantics contain restrictions on the ghost state that capture the
invariants we want to prove for the back-translated program.
For instance, below one can see what we added to the rules to capture the well-bracketedness of
the stack: calls add frames to the call stack, and returns are only allowed when
the top frame of the stack matches that return.
\[
\inferrule[S-Step-Stack-Call]{
  \dots \\
  \mathsf{gs'}.\mathrm{stack} = (\prg{C_i}, \prg{C_j}) :: \mathsf{gs}.\mathrm{stack}
}{ s, \mathsf{gs} \xrightarrow{\ecall{\prg{C_i}}{\prg{C_j}}{\prg{z}}} s', \mathsf{gs'} }
\\
\inferrule[S-Step-Stack-Return]{
  \dots \\
  \mathsf{gs}.\mathrm{stack} = (\prg{C_j}, \prg{C_i}) :: \mathsf{s} \\
  \mathsf{gs'}.\mathrm{stack} = \mathsf{s}
}{ s, \mathsf{gs} \xrightarrow{\eret{\prg{C_i}}{\prg{C_j}}{\prg{z}}} s', \mathsf{gs'} }
\]

This ghost state is an abstraction of the concrete state of the generated source code.
For instance, the abstract location stored in trees correspond to
the private variable \(\loc\) of each generated compartment,
and the abstract stack corresponds to the concrete stack of the source
language in the sense that it captures the same cross-compartment behaviors.

We build the semantics iteratively, by taking the semantics of the
previous level, and adding the new conditions in addition of the previous ones.

\paragraph{Initialization}
We do not start from trees, but actually from sets of traces that satisfy a notion of
well-formedness.
For this reason, the first step of our back-translation is to constructively prove that such sets
of traces can be represented as trees satisfying the predicates described previously.
To do so, we define \lstinline[language=Rocq]{add_trace_to_tree}
(\autoref{lst:initialization-functions}) that adds a trace to a tree,
and \lstinline[language=Rocq]{tree_of_trace_list} that iterates this function
over a sequence of traces.
\begin{lstlisting}[numbers=left,language=Rocq,keepspaces=true,basicstyle=\footnotesize,
xleftmargin=1.5\parindent,
caption={Construction of trees from traces},label=lst:initialization-functions]
Fixpoint add_trace_to_tree (tra: trace) (tr: tree unit): tree unit :=
  match tra with
  | nil => tr
  | cons e q =>
      match tr with
      | node tt br => node tt (add_trace_to_branch e q br)
      end
  end
with add_trace_to_branch (e: event) (tra: trace) (br: branches unit): tree unit :=
  match br with
  | Bnil => branch_of_trace (e :: tra)
  | Bcons e' tr br' => if (e == e') then
                        Bcons e (add_trace_to_tree tra tr) br'
                      else
                        Bcons e' tr (add_trace_to_branch e tra br')
  end.
\end{lstlisting}
The core of this function is
actually \lstinline[language=Rocq]{add_trace_to_branch}: this function takes an
event and a trace following this event, traverses the branches until it finds
one that is already labeled by this event, and merges the remaining trace with
that branch (line 12-13).
If it doesn't find such a branch, it creates a new one (line 11).
This ensures there is no duplication between events, as required by \lstinline[language=Rocq]{deterministic_tree}.

We do not reproduce our notion of well-formedness here, but it captures the determinacy
conditions for proving \lstinline[language=Rocq]{unique_current_tree}: that is,
if the history is the same and the context has control, then it can't be that
two different events appear at this point on the traces.

\paragraph{Tree transformations}
Steps (b) and (c) are simple tree transformations. Pass (b)
adds a unique identifier and pass (c) a stack snapshot to each node.%
The stack snapshot records what the state is supposed to be when reaching that node, which simplifies the statement of the properties
and the reasoning in the simulation proof.
These two transformations compile trees by adding new information to the nodes of the trees:
pass (b) goes from \lstinline[language=Rocq]{tree unit} (level 1) to \lstinline[language=Rocq]{tree nat} (level 2); and pass (c) goes from this to \lstinline[language=Rocq]{tree (nat * stack)} where \lstinline[language=Rocq]{stack = list (Compartment.id * Compartment.id)}.
These transformations are simple (they map a transformation over all nodes), and
do preserve the determinacy invariants
described previously. The more involved transformations, are
described in the next subsection.
\subsection{Flattening and code generation}\label{sec:flattened}
The trees used so far cannot be
easily mapped to source compartments in one go, as described in \autoref{sec:back-by-example}.
Instead, we go through two steps, (d) and (e), that manipulate a flattened representation
of the trees, in order to generate source statements.

\paragraph{Flattening}
Step (d) flattens trees into a list of rules
describing how compartments should update their location when they get
control, and what event they need to produce at a given location.
Concretely, a program at level (4) is a map from compartments to
\lstinline[language=Rocq]{list (nat * event * nat)}.
Elements of such a list are rules on how to update location and what events to emit.
As an additional invariant, these lists of rules satisfy uniqueness and determinism properties that
derive from the uniqueness and determinism properties of the tree languages.
For example, consider a call
\(\ecall{\prg{C_1}}{\prg{C_2}}{40}\)
that appears in a trace.
Then:
\begin{itemize}
\item the list corresponding to \(\prg{C_1}\) will contain some rule
\((n_1, \ecall{\prg{C_1}}{\prg{C_2}}{40}, n_1')\) meaning that when \(\prg{C_1}\) is in
location \(n_1\), then it must update its location to \(n_1'\) and call
\(\prg{C_2}\) with argument \(40\). Additionally, this list will not contain any other
rule starting at \(n_1\) (because \(\prg{C_1}\) has control, it can only do one next event);
\item the list corresponding to \(\prg{C_2}\) will contain some rule
\((n_2, \ecall{\prg{C_1}}{\prg{C_2}}{40}, n_2')\) meaning that when \(\prg{C_2}\) is called
with argument \(40\) and its current location is \(n_2\),
it must update its location to \(n_2'\). Additionally, the list may contain other rules
starting at \(n_2\), however none of those will also be for a call with the specific argument \(40\).
\end{itemize}
We give a semantics to this language in the same style as for the trees.
The main difference is that a step can be taken iff a corresponding rule appears
in the mapping for each compartment.
Unlike the tree representation, the list of rules is \emph{not} reduced during execution:
this is similar to the source semantics in that the source code of a source compartment
do not ``disappear'' when it is executed.
Finally, the ghost state still keeps track of the abstract call stack.

\paragraph{Code generation}
The code generation step (e) transforms the previous lists of rules to executable source code.
For each rule, this step generates two statements: one for the compartment giving up control,
and the other for the one gaining control.
The statements of each compartment are then concatenated so as to obtain source
code akin to \autoref{fig:ex-tree}, by filtering and folding over the list of
rules and statements.

\section{Proving the back-translation correct using forward simulations}\label{sec:proof-sim}
In this section, we discuss the core of our proof: how we prove forward
simulations between all the tree languages to obtain a correctness lemma
for each of these steps.

\subsection{Forward simulations}
To prove \autoref{lemma:correct-step} between tree languages, we
use \emph{forward simulations} in the style of CompCert.
For each step \(\alpha\) between levels \((i)\) and \((i+1)\), we define a
simulation relation \(\sim_{(\alpha)}\) between the states of two consecutive languages.
Then, we prove the following lemmas:
\begin{lemma}[Initial states are related]\label{lemma:fwd-sim-ini}
Let \(s_i\) be an \emph{initial} state at level \(i\) and \(s_{i+1}\) an \emph{initial} state at level (i+1),
then \(s_i \sim_{(\alpha)} s_{i+1}\).
\end{lemma}

\begin{lemma}[Forward simulation]\label{lemma:fwd-sim}
  Let \(s_i\) be a state at level i and \(s_{i+1}\) a state at level (i+1).\\
  Suppose \(s_i \sim_{(\alpha)} s_{i+1}\) and \(s_i \xrightarrow{e}_i s_i'\).
  Then there exists a state \(s_{i+1}'\) at level (i+1) such that:
  \begin{equation*}
  \begin{split}
  s_{i+1} \xrightarrow{e}_{i+1}^* s_{i+1}'
  \end{split}
  \quad\quad\text{and}\quad\quad
  \begin{split}
  s_i' \sim_{\alpha} s_{i+1}'
  \end{split}
  \end{equation*}
\end{lemma}
It is well known~\cite{Leroy09b,TanMKFON19} that, together, these two lemmas imply \autoref{lemma:correct-step}.

\subsection{Simulation relations and proofs}

We now describe how to define these simulation relations for each pass.

\paragraph{Simulations between tree languages}
Recall that for the tree languages, states contain a map from compartment identifier to trees
storing what is left to be executed.
For passes between such tree languages, the basic idea is simple, and can be
summarized as follows: we always maintain one essential property: if \(s_i\) is a state at level \(i\) and \(s_{i+1}\) a state at level \(i+1\) that ought to be
related by the simulation relation, then for any compartment \(\prg{C_i}\),
\(s_{i+1}.\mathrm{trees}(\prg{C_i}) = f (s_i.\mathrm{trees}(\prg{C_i}))\)
where \(f\) is the transformation applied by the pass.
In addition, when we introduce the abstract stack, the simulation relation captures
the fact that this stack is always well-bracketed.
We use the predicate in \autoref{lst:wf-stack-trace} in the simulation relation between the states at level \(2\) and \(3\):
\begin{lstlisting}[numbers=left,language=Rocq,keepspaces=true,
basicstyle=\footnotesize,xleftmargin=1.5\parindent,
caption={Relation for the stack used in the simulation proof between levels \(2\) and \(3\)},label=lst:wf-stack-trace]
Fixpoint wf_stack_trace (tra: trace) (st: stack): bool :=
  match tra with
  | [] => true
  | ECall Ccur P z Cnext :: tra' => wf_stack_trace tra' ((Ccur, P, Cnext) :: st)
  | ERet Ccur z Cnext :: tra' =>
      match st with
      | (Cnext', _, Ccur') :: st' =>
          (Ccur == Ccur') && (Cnext == Cnext') && wf_stack_trace tra' st'
      | _ => false
      end
  end.
\end{lstlisting}
This predicate relates the trace remaining to be executed and the current stack: if the next event is a call (line 4), then the rest of the trace is related to the same stack, with a new frame corresponding to the call added.
On the other hand, if the next event is a return (line 5), then the
stack must have a top stackframe that corresponds to this return and the
rest of the trace must be related to the rest of the stack.
We use this predicate in the simulation to prove the well-bracketedness of the stack required by the semantics.

Note that we don't need to manipulate this predicate after step (c),
because after that the semantics of the languages enforce the well-bracketedness of the stack, as in \autoref{sec:tree-langs-struct}.

\paragraph{Flattening}
Recall that after flattening, we do not reduce the list of rules anymore; hence,
the list of rules in the flattened representation is a superset of the trees from
the previous level.
Instead, the invariant we maintain states the following: any rule that can be
derived from the source trees in the current state can be found in the list of
rules we keep.

In addition, we prove the determinacy invariants stated
in \autoref{sec:flattened} explicitly from the start,
and make it part of a notion of well-formedness of our programs.
We use this to prove the notion of uniqueness from \autoref{lst:uniqueness-flat}:
\begin{lstlisting}[numbers=left,language=Rocq,keepspaces=true,basicstyle=\footnotesize,
xleftmargin=1.5\parindent,
caption={Uniqueness for the flattened representation},label=lst:uniqueness-flat]
Definition uniqueness_in {A B: Type} (l: list (B * A)) :=
  forall b a a', In (b, a) l -> In (b, a') l -> a = a'.

Definition unique_incoming_calls: forall C es,
  prog_rules p C = Some es ->
  uniqueness_in (incoming_calls C es).
\end{lstlisting}
This says that in the list of incoming calls for compartment \(\mathtt{C}\),
there is at most one rule for each key, and here keys are of type
\lstinline[language=Rocq]{(nat * Z)}: the natural number corresponding to the
location, and the integer argument for this rule.
We have similar definitions for incoming returns and for outgoing events.
Proving that the properties of the trees imply these uniqueness properties is quite
involved, but only because of a complex induction on trees.

\paragraph{Code generation}
Step (e) goes from the flattened representation to source code
and could have been the most complex to reason about.
However, since the flattened representation is very close in its structure to the actual
code produced by the back-translation, most of the reasoning
to prove this step correct has already been done.
This proof step goes as follows:
\begin{enumerate}
\item We define the simulation invariant that hold at each synchronization point
  of the simulation (\IE{} when we have to prove the two states are
  related).%
\item We prove individual properties of the back-translations of each rule: they produce the right event, they update the location correctly, \ETC{}
\item We prove that the generated code chooses the right branches in the branching structure, using
the determinism and uniqueness properties of the flattened representation.
\item We assemble the full programs, and prove the forward simulations, using all the above properties and
the determinism and uniqueness properties of the flattened representation.
\end{enumerate}
The core of our simulation invariants at reproduced in \autoref{lst:invariant} below:
\begin{lstlisting}[numbers=left,language=Rocq,keepspaces=true,%
basicstyle=\footnotesize,
xleftmargin=1.5\parindent,
caption={Flattened to source simulation invariant},label=lst:invariant]
Inductive match_concrete_stacks: Component.id -> stack -> CS.stack -> Prop :=
| match_concrete_stacks_nil: forall C, match_concrete_stacks C [] []
| match_concrete_stacks_int: forall C s st old_arg,
    match_concrete_stacks C s st ->
    match_concrete_stacks C s (CS.Frame C old_arg Kstop :: st)
| match_concrete_stacks_ext: forall C C' s st old_arg,
    C <> C' ->
    match_concrete_stacks C' s st ->
    match_concrete_stacks C
      ((C', C) :: s)
      (CS.Frame C' old_arg (Kafter_call C' 0) :: st) .

Definition match_mem (locs: NMap nat) (mem: Memory.t) :=
  forall C n, locs C = Some n ->
    Memory.load mem (C, Block.local, 0%
    /\ Memory.load mem (C, Block.local, 1%
    /\ (exists v, Memory.load mem (C, Block.local, 2%

Inductive match_states (p: Flattened.prg) (i: nat): Flattened.state -> CS.state -> Prop :=
| match_states_cons: forall t locs st C cst mem e v es,
    forall (MS_ST: match_concrete_stacks C st cst)
      (MS_MEM: match_mem locs mem)
      (ES: Flattened.genv_code (globalenv (Flattened.sem p)) C = Some es),
      e = (switch_outgoing C es E_exit) ->
      match_states p i (t, locs, st) (CS.State C cst mem Kstop e v) .
\end{lstlisting}
This simulation invariant states that:
two states are related if the concrete program is currently
executing a compartment's code (line 23-24: \lstinline[language=Rocq]{switch_outgoing} is a function that unfolds to the code displayed in the second half of the code in \autoref{fig:ex-tree});%
the abstract and the concrete stack are related;
and the memory in the source state is related to the map of locations in the
flattened %
state.
The stacks are related (\lstinline[language=Rocq]{match_concrete_stacks}) when they represent the
same cross-compartment call-stack. The rule~\lstinline[language=Rocq]{match_concrete_stacks_int} allows the source to perform internal calls, which are then unfolded during the simulation proof when returning.
The memory must agree with the location map (line 15), and the other variables must be accessible.

While the proof of this pass is the longest in our development, it consists mostly in
step-by-step reasoning about the behavior of source statements, which is made much easier
because all the harder invariants have already been established in previous steps.

\section{Formalization in Rocq}\label{sec:implementation}
We have implemented and verified this back-translation in the
Rocq proof assistant.
Starting from a well-formed set of traces, we define each step of the
back-translation from (a) to (e) and the semantics of each intermediate
language. We prove an initialization lemma converting well-formed sets of traces to trees,
and for each remaining step we prove a CompCert-style forward simulation result,
following the ideas described in the previous sections.
Our Rocq development is, however, done in a slightly more complex
setting~\cite{AbateABEFHLPST18} than the simplified one we used for aiding
the exposition above.
In the Rocq development we consider multiple procedures per compartment, that are
mediated through an interface, which is interesting, since the compartments are
able to distinguish when receiving calls to different procedures.
Therefore, we must generate slightly different statements for handling calls
for each procedure of a compartment.
We do this by recording the procedure being called in the call events.

The development is %
$\sim$7800 lines of Rocq code long (comments
excluded), with 1200 lines of specifications and 6600 of proofs. This figure
includes the definition of the source language, but neither that of the trace model,
nor that of some libraries and existing files that we reuse from Abate \ETAL~\cite{AbateABEFHLPST18}'s
development.
The largest part of the development is spent reasoning about the determinacy invariants, in particular proving that they imply the uniqueness from \autoref{lst:uniqueness-flat}.
This part should be fully reusable with minor adaptation, because it takes place
between levels \(3\) and \(4\), before we start reasoning about the source
language.

\paragraph{Benefits of our approach}
We found approaching the proof in nanopass style beneficial, since it
allowed us to partition proof concerns relatively easily.
In particular, we were able to parallelize the work between the first steps and the
latter steps without too much effort, because these steps were not directly related.
On the contrary, when we previously attempted a monolithic proof, any
modification at any level of the proof did affect the entire development and made
the verification of the many invariant complex and more time-consuming.

\section{Related Work}\label{sec:related}

The kind of traces used for trace-based back-translation come from previous work
on ``fully-abstract trace semantics''~\cite{JeffreyR05, JeffreyR05b, Laird07}, a
form of game semantics~\cite{AbramskyJM00, HylandO00, GhicaT12}.
Trace-based back-translation was extensively investigated for proving fully
abstract compilation~\cite{PatrignaniC15, BusiNBGDMP20, akram-capabileptrs, AgtenSJP12}.
In that setting, one needs to back-translate two trace prefixes that differ only in one
single final event, which is much simpler than back-translating a finite set of
trace prefixes.

We instead build on the work of Abate \ETAL~\cite{AbateABEFHLPST18}, who prove robust
safety preservation for a stateful source language.
As discussed in \autoref{sec:background}, their secure compilation proof can be
adapted to our setting, with the notable exception of the back-translation step.
Their secure compilation criterion only requires back-translating a single trace
prefix, which is again much easier than a finite set of trace prefixes.
Another work by Abate \ETAL~\cite{AbateBGHPT19} does look at finite sets of trace prefixes,
but for a much simpler source language with neither state nor callbacks, which
are the features that cause the biggest challenges for our back-translation.
Yet even in that simplified setting their back-translation does get reasonably
complicated and their proofs are done only informally on paper.
We believe that formalizing those proofs in a proof assistant could
greatly benefit from our nanopass approach.

Trace-based back-translation is not the only way to do back-translation.
Especially for proving fully abstract compilation, context-based
back-translations are also quite popular, \IE back-translations that only look
at (the syntax of) the target context for producing a source context.
The context-based back-translation techniques used for fully abstract compilation
include: 
embedded interpreters~\cite{FournetSCDSL13},
universal embedding~\cite{NewBA16},
approximate back-translation~\cite{DevriesePPK17},
universal contracts~\cite{StrydonckPD19}, \ETC
Abate \ETAL~\cite{AbateBGHPT19} show that such context-based back-translations can be used
to prove stronger secure compilation criteria, although the practical need
for some of the strongest criteria is still unclear.
We believe that robust hypersafety preservation~\cite{AbateBGHPT19} is already
good enough for most practical applications, and the only reason we prove the
stronger \RFrSP{} is to show that we can do it without any significant increase
in the complexity of the back-translation.
A more practically-relevant advantage of context-based back-translations is that
they can usually deal with higher-order languages.
One disadvantage we see is that context-based back-translations are often
very specific to the details of the source and target languages involved, while
for trace-based back-translation the traces can abstract away unnecessary
details, which we think can lead to better reusability across languages.
Finally, context-based back-translation does not seem applicable when the source
and target of the compiler are very different: for the languages we inherited
from Abate \ETAL~\cite{AbateBGHPT19} the source has structured control flow, while the
target language does not.

Only a few of the existing back-translations have been
formalized in a proof assistant~\cite{DevriesePPK17, AbateABEFHLPST18,
ThibaultBLAAGHT24, secureptrs, DevrieseMP24}.
Secure compilation is, however, a larger area, and not all notions of secure
compilation rely on back-translation.
For instance, Besson \ETAL~\cite{BessonBDJW19} introduce a software-fault isolation pass
to one of the early intermediate languages of CompCert.
This pass adds masking and checks to turn undefined behavior into defined
behavior that only affects the component causing it.
Other notions of secure compilation focus not on protecting trusted from
untrusted code, but on preserving security (hyper)properties such as
information flow control~\cite{BessonDJ19} and cryptographic constant-time~\cite{AlmeidaBBBGLOPS17,BartheGL18,barthe-popl20}.
For these works on cryptographic constant-time, they prove their results without back-translation,%
by augmenting the semantics with notions of
leakage and proving the preservation of leakage equivalence.

\section{Future Work}\label{sec:future}

In the future, we plan to reuse our proof technique for other languages and
compilation chains, and to scale it up to formally proving the stronger \RFrSP
security criterion for the compartmentalized CompCert variant recently proposed
by Thibault \ETAL~\cite{ThibaultBLAAGHT24}.
This variant aims at providing a safer semantics to C, in which the damage from
undefined behavior is restricted to only compromising the compartment that
explicitly triggered it~\cite{AbateABEFHLPST18}, \EG by writing out of the
bounds of an array.
Since Thibault \ETAL~\cite{ThibaultBLAAGHT24} only prove a variant \RSP, strengthening this to
\RFrSP could greatly benefit from our nanopass back-translation technique.
Moreover, the first steps of our nanopass back-translation are independent of the details
of the source language, and only rely on properties of the trace model.
Moreover, even the code-generating step (e) should be reusable for other languages,
since it relies on a limited set of standard language features, such as
procedures and private state where compartments can store the location
information.
Hence, we believe that our back-translation technique will scale up
to the more realistic setting of this CompCert variant.

Finally, we would also like to extend nanopass back-translation to the fine-grained
memory sharing that occurs when passing secure pointers to
data~\cite{akram-capabileptrs} and code %
between compartments.

\clearpage
\bibliographystyle{plainurl}
\bibliography{refs,safe,mp}
\end{document}

%% file: abstract.txt
Researchers aim to build secure compilation chains enforcing that if there is no
attack a source context can mount against a source program then there is also no
attack an adversarial target context can mount against the compiled program.
Proving that these compilation chains are secure is, however, challenging, and
involves a non-trivial back-translation step: for any attack a target context
mounts against the compiled program one has to exhibit a source context mounting
the same attack against the source program. We describe a novel back-translation
technique, which results in simpler proofs that can be more easily
mechanized in a proof assistant. Given a finite set of finite trace prefixes,
capturing the interaction recorded during an attack between a target context and
the compiled program, we build a call-return tree that we back-translate into a
source context producing the same trace prefixes. We use state in the generated
source context to record the current location in the call-return tree. The
back-translation is done in several small steps, each adding to the tree new
information describing how the location should change depending on how the
context regains control. To prove this back-translation correct we give
semantics to every intermediate call-return tree language, using ghost state to
store information and explicitly enforce execution invariants. We prove several
small forward simulations, basically seeing the back-translation as a verified
nanopass compiler. Thanks to this modular structure, we are able to mechanize
this complex back-translation and its correctness proof in the Rocq prover
without too much effort.



%% file: img/technique.tex
\begin{tikzpicture}[auto]
  \node(cPi)
       {\(\red{\forall i.}~\cmp{\mathbb{P}_{\red{i}}} \cup \mathbb{C}_{\mathtt{T}} \rightsquigarrow m_{\red{i}} \)};

  \node[right = of cPi] (CP1t)
       {\(\red{\forall i.}~\cmp{\mathbb{P}_{\red{i}}'} \cup \cmp{\mathbb{C}_{\mathtt{S}}} \rightsquigarrow m_{\red{i}} \)};

  \node[align = left, above = of cPi] (CP1s)
       { \(\back{\red{(m_1,\dots,m_K)}}~ = (\mathbb{C}_{\mathtt{S}}, \mathbb{P}_{\red{1}}'\red{, \dots \mathbb{P}_K'})\) \\
    \(\red{\forall i.}~\mathbb{P}_{\red{i}}' \cup \mathbb{C}_{\mathtt{S}} \rightsquigarrow m_{\red{i}}\)};

  \node[right = of CP1t, xshift=2em] (CPt)
       {\(\red{\forall i.}~\cmp{\mathbb{P}_{\red{i}}} \cup \cmp{\mathbb{C}_{\mathtt{S}}} \rightsquigarrow m_{\red{i}} \)};

  \node[above = of CPt] (CPs)
       {\(\red{\forall i.}~\mathbb{P}_{\red{i}}\cup\mathbb{C}_{\mathtt{S}} \rightsquigarrow m_{\red{i}}\)};

  \draw[->] (cPi.90) to node {\em I Back-translation} (CP1s.-90);

  \draw[->] (CP1s.-45) to node [align=left,font=\itshape,xshift=-2em,yshift=0.75em]{II Forward Compiler Correctness} (CP1t.180);

  \draw[->] (CP1t.0) to node [below,xshift=0.5em,yshift=-0.5em]{\em III Recomposition} (CPt.180);
  \coordinate [below=1.5em of CP1t,xshift=0em] (compoint1);
  \coordinate [below=1.5em of CPt,xshift=-4em] (compoint2);
  \draw[-] (cPi) to (compoint1);
  \draw[-] (compoint1) to (compoint2);
  \draw[->] (compoint2) to (CPt.220);

  \draw[->] (CPt.90) to node [left,align=left,font=\itshape,yshift=-0.5em]{IV Backward Compiler Correctness} (CPs.-90);

\end{tikzpicture}